\documentstyle[12pt]{article}
\title{ Determination of Dipole Polarization Effects in $^7$Li and $^{11}$Li}

\author{L.D.~Skouras\thanks{Permanent address: Institute of Nuclear Physics,
 NCSR {\it Demokritos}, GR--15310 Aghia Paraskevi, Attiki, Greece} and
 H.~M\"{u}ther\\
Institut f\"{u}r Theoretische Physik der Universit\"{a}t T\"{u}bingen\\
D-7400 T\"{u}bingen, Federal Republic of Germany\\[5mm]
   and\\[5mm]
   M.A.~Nagarajan\\
     SERC Daresbury Laboratory\\
     Daresbury, Warrington WA4 4AD, UK}
\date{}

\begin{document}
\baselineskip=18pt plus 1pt minus 1pt
\lineskip=18pt plus 1pt minus 1pt
\setcounter{equation}{0}

\maketitle

\vspace{10mm}
\begin{abstract}
\noindent
The structure of $^6$Li, $^7$Li and $^{11}$Li nuclei is investigated in a
model space which includes all configurations with oscillator energy up to
$3\hbar\omega$ above the ground state configuration. The energy spectra and
electromagnetic properties of the low-lying states are determined with various
two-body interactions, which are derived from the Bonn potential. The results
of these shell-model calculations depend on the strength of the tensor
component contained in the NN interaction and also on the treatment of the
Dirac spinors for the nucleons in the nuclear medium. In addition
the calculation determines the {\it dipole polarizability} of $^7$Li and
$^{11}$Li caused from virtual $E1$ excitations to the positive parity states
of these nuclei. It is demonstrated that the BAGEL scheme provides a very
powerful tool to consider contributions from virtual excitations up to large
energies.
\end{abstract}

\vfil\eject

\section{Introduction}

The static and dynamic electric moments of nuclei are traditionally measured
by Coulomb excitation$^1$). The quadrupole moments of the excited nuclear
states can be measured by the reorientation method$^2$). It has long been
established that the quadrupole moments as well as the $B(E2)$ values can
be affected from virtual excitations to the giant dipole state$^3$). In
particular, if the dipole state occurs at low excitation energy, it can cause
a measurable renormalization of the $E2$ operator.

The most careful measurements of the correction due to the dynamic {\it dipole
polarizability} on the $B(E2)$ and the quadrupole moments has been in the case
of the nucleus $^7$Li~$^{4-5}$). This nucleus has a ground state with
$J^{\pi} = 3/2^-$ and a first excited state with $J^{\pi} = 1/2^-$.
The excited state has no spectroscopic quadrupole moment and hence a precise
determination of the $B(E2)\uparrow$ value for the excitation of
the $1/2^-$ state
over a range of energies will enable one to measure the {\it inelastic
dipole polarizability}. Similarly, a careful measurement of the ground
state quadrupole moment will enable one to measure the {\it elastic
dipole polarizability}. The results of various measurements are reviewed
by Barker {\it et al}~$^6$) and Voelk and Fick$^7$).

There exist several theoretical investigations on the dipole polarization
effects of $^7$Li. Several of these calculations adopt the cluster-model
approach$^{8-11}$). On the other hand the shell-model approach has been
followed by Gomez-Comacho and Nagarajan$^{12}$) and Barker and Woods$^{13}$).
In both these calculations the natural parity states are determined considering
only $0p$ shell configurations while the unnatural parity states are studied in
the space of the $1\hbar\omega$ excitations.

Recent experiments$^{14}$) suggest that a component of the dipole resonance,
termed {\it the soft dipole mode}$^{15}$) or {\it the pygmy resonance}$^{16}$),
occurs at a very low excitation in $^{11}$Li. This feature combined with the
large $r_{ms}$ radius of $3.16 \pm 0.11 fm$~$^{17}$) have made
this nucleus the subject of several theoretical investigations$^{18-28}$). In
these studies various approaches have been adopted, like the cluster-
model$^{18-21}$), the random phase approximation$^{22}$), the sum-rule
approach$^{23}$) and the mean-field description$^{24-25}$). In addition the
structure of $^{11}$Li has been investigated in two recent large scale
shell-model calculations by Hoshino {\it et al}~$^{26}$) and Hayes and
Strottman$^{27}$). In both these calculations, the properties of the natural
parity states are described in the space of $(0+2)\hbar\omega$ excitations.
For the unnatural parity states these authors employ a model space,
which contains the $1\hbar\omega$ configurations and some selected
$3\hbar\omega$ excitations. A common
feature in these two shell-model calculations is the manner in which the
effective hamiltonian is chosen. This consists of a mixture of hamiltonians
each of which was originally constructed for calculations in smaller model
spaces. Thus Hayes and Strottman$^{27}$) use the Cohen-Kurath$^{28}$)
p-shell matrix elements, the Millener-Kurath$^{29}$) particle-hole interaction
and single particle energies and the Chung$^{30}$) sd-shell matrix elements. A
similar approach is also adopted in the calculation of Hoshino {\it et
al}~$^{26}$).

In this paper we present a shell-model study of the natural parity states
of $^6$Li, $^7$Li and $^{11}$Li. The properties of these states are
investigated in the space of $(0+2)\hbar\omega$ excitations.
For $^7$Li and $^{11}$Li we also examine the unnatural parity states for which
we consider the complete space of $(1+3)\hbar\omega$ excitations. The most
detailed data exist for $^7$Li and we use them to draw conclusions about
the reliability of our model. On the other hand, there exist very little
data for $^{11}$Li and thus our results should be taken as theoretical
predictions.

The main purpose of our investigation has been to determine the $E1$
polarization effects on the quadrupole moment of the $3/2^-$ ground state of
both $^7$Li and $^{11}$Li as well as on the $B(E2)\uparrow$ value corresponding
to the excitation of the  $1/2^-$ state of $^7$Li. Several sets of results
have been determined corresponding to different hamiltonians. To test the
reliability of these calculations we have also determined the properties of
the low-lying states of $^6$Li, $^7$Li and $^{11}$Li and compare the
theoretical predictions with the existing experimental data.

The present calculation differs from previous related studies of
$^7$Li~$^{12-13}$) and $^{11}$Li~$^{26-27}$) in several aspects.
One of them is the use of a larger model space and of a realistic hamiltonian
which contains no adjustable parameters. Thus in our calculation we consider
harmonic oscillator energies and wavefunctions for the single-particle states,
while for the two-body part of the hamiltonian we employ $G$-matrices that
have been derived from the Bonn potential$^{31}$). Since our model space,
although large, is restricted, the final results depend on the choice of the
harmonic oscillator size parameter.
We determine an optimal oscillator parameter by requiring that the
calculated energy for the ground state for the nucleus is a minimum with
respect to a change of the oscillator basis. Another new feature of our
calculation is that we take into account a possible change of the Dirac
spinor for the nucleons in the nuclear medium as compared to the Dirac
spinors in the vacuum. Recent investigations on open shell nuclei$^{32,33}$)
have demonstrated that effects of a realistic spin-orbit
splitting can only be obtained if this relativistic
feature of nuclear many-body theory is taken into account.
Finally, we consider bare electromagnetic operators. We expect the
renormalization
of the operators to arise naturally from the use of the large model space.
By considering different model spaces and their effect on the predictions
of the calculation we expect to obtain a better understanding of the
relationship between our approach and the cluster model$^{8-11,18-21}$).

In sect. 2 we discuss the details of our calculation, while in sect. 3 we
present our results. Finally, sect. 4 contains the conclusions of this work.

\section{Details of the calculation}
\subsection{\sc Definitions}

The effective potential which is used to describe $E2$ excitation of the
projectile in Coulomb scattering experiments consists of two parts$^1$) :
\begin{equation}
\label{df1}
V_{E2} =V_{coup} +V_{pol} \quad ,
\end{equation}
where
\begin{equation}
\label{df2}
\langle f\vert V_{coup}(\vec{r})\vert i\rangle ={{4\pi}\over 5}
{{Ze}\over r^3}
\langle J_f \vert \vert E2\vert \vert  J_i\rangle
X^{J_fJ_i}_{M_fM_i}(\hat{r}) \quad ,
\end{equation}
\begin{equation}
\label{df3}
\langle f\vert V_{pol}(\vec{r})\vert i\rangle
= -{{Z^2e^2}\over r^4}[\delta_{if}P
+\sqrt {{9\pi}\over 5} (-1)^{J_f-J_i}
\tau _{if} X^{J_fJ_i}_{M_fM_i}(\hat{r})] \quad ,
\end{equation}
with
\begin{equation}
\label{df4}
X^{J_fJ_i}_{M_fM_i}(\hat{r}) = {1\over (2J_f+1)^{1/2}} \sum_\mu
Y^\ast _{2\mu} (\hat{r}) \langle J_i M_i 2 \mu \vert  J_f M_f\rangle
\end{equation}

In the above eqs (\ref{df2})-(\ref{df4}), $Z$ is the charge
number of the target, $\vec{r}$ is the distance between the
two nuclei, while $\vert f\rangle$ and $\vert i\rangle$ denote the
final and initial states of the projectile, respectively.

The polarization potential (\ref{df3}) consists of a monopole and a
quadrupole term. The monopole term is defined$^1$) as:
\begin{equation}
\label{df5}
P={{4\pi} \over 9}\sum _n {B(E1; i\rightarrow n) \over (E_n -E_i)}
\end{equation}
where
\begin{equation}
\label{df6}
B(E\lambda ; i\rightarrow f) = {\langle f\vert \vert E\lambda
\vert \vert i\rangle ^2\over
2J_i+1}
\end{equation}
and $\vert n\rangle$ are the states which are connected by the
$E1$ operator with
the state $\vert i\rangle$.

The quantity $\tau _{if}$ in the quadrupole part of $V_{pol}$ is the tensor
moment of the electric polarizability and is defined$^{4,34}$) as:
\begin{equation}
\label{df7}
\tau_{if} = {{8\pi} \over 9} \sqrt {10 \over 3} \sum _n W(11J_fJ_i;2J_n)
{{\langle i\vert \vert E1\vert \vert n \rangle \langle n \vert
\vert E1\vert \vert  f \rangle }\over (E_n -E_i)}
\end{equation}
The relation between $\tau _{if}$ and the quantity $S(E1)$ defined by
H\"{a}usser {\it et al}~$^{35}$) is
\begin{equation}
\label{df8}
\tau _{if}= {{8\pi} \over 9} \sqrt {10 \over 3} S(E1)
\end{equation}

{}From the analyses of the $^7$Li data$^{6,7}$) values for four quantities
have been extracted. These are
$B(E2;1\rightarrow 2)$, $\tau_{12}$, $\tau_{11}$ and the
static quadrupole moment $Q_s$. We use here the notation usually
adopted$^{6,7}$) where 1 refers to the ground state and 2 to the first excited
state. $Q_s$ is defined as:
\begin{equation}
\label{df9}
Q_s = \sqrt{{16\pi} \over 5}\sqrt{{J_i(2J_i-1)}
\over {(J_i+1)(2J_i+1)(2J_i+3)}}\langle i \vert \vert E2\vert
\vert i\rangle \quad .
\end{equation}
The experimental values for these four quantities are compared with the
predictions of the calculation in sect. 3.

\subsection{\sc The shell-model calculation}

In this section we discuss the details of the shell-model calculation on
the Li isotopes.

We consider a hamiltonian of the form
\begin{equation}
\label{sm1}
H=\sum _i ^A t_i + \sum _{i<j}^A G_{ij} + {1\over 2} A m \omega ^2 R ^2
-{3\over 2}\hbar\omega \quad ,
\end{equation}
where $t$ denotes kinetic energy, $G$ is the two-body interaction while $A$
and $\vec R$ stand for the mass number and center-of-mass coordinate of the
nucleus,
respectively. The term ${1\over 2} Am\omega ^2R^2$ has been included in the
hamiltonian $H$ in order to remove spurious state effects$^{36}$) while the
term $-{3\over 2}\hbar\omega$ removes the center-of-mass contribution from
a non-spurious state. The hamiltonian (\ref{sm1}) can be written as:
\begin{equation}
\label{sm2}
H=H_0 + V \quad ,
\end{equation}
where
\begin{equation}
\label{sm3}
H_0 = \sum _i ^A (t_i +{1\over 2} m\omega ^2 r_i^2) -{3\over 2}\hbar\omega
\; , \quad\quad V= \sum _{i<j} ^A (G_{ij} -{{m\omega ^2}\over {2A}}
(\vec{r}_i -\vec{r}_j)^2) \quad .
\end{equation}

The basis of our calculation consists of all eigenvectors of $H_0$
which have unperturbed energy not exceeding by $3\hbar\omega$ the energy
of the ground state configuration $(os)^4(op)^{A-4}$. Thus the model space
contains 15 single particle orbitals from the $0s$ up to the $sdg$ shell
of the harmonic oscillator potential. The basis vectors
are constructed assuming no core state and are represented as
\begin{equation}
\label{sm4}
\vert \Phi\rangle =\vert C^A; JT \mu \rangle \quad ,
\end{equation}
where $C^A$ denotes a distribution of the $A$ particles among the single
particle-orbitals, while the index $\mu$ distinguishes the vectors which
correspond to the same set of $\{ C^A,J,T \}$ quantum numbers. Matrix elements
of operators between the states (\ref{sm4}) have been determined using
the {\it generalized fractional parentage} formalism of Skouras and
Kosionides$^{37}$).

Since the vectors (\ref{sm4}) have been constructed using isospin formalism
it is convenient to consider in the same formalism the operators $E\lambda$
of sect. 2.1. Thus
\begin{equation}
\label{sm5}
E^\lambda _\mu = E^{\lambda 0}_{\mu 0} + E^{\lambda 1}_{\mu 0} \quad ,
\end{equation}
where
\begin{equation}
\label{sm6}
E^{\lambda 0}_{\mu 0} = {e\over 2} \sum _i ^ A r^\lambda _i
Y _{\lambda \mu}(\hat{r}_i) \; , \quad
E^{\lambda 1}_{\mu 0} = {e\over 2} \sum _i ^ A \tau _0(i) r^\lambda _i
Y _{\lambda \mu}(\hat{r}_i) \quad .
\end{equation}
Thus the matrix elements of the operator $E\lambda$ can be expressed as:
\begin{eqnarray}
\label{sm7}
\langle J_f T_f M_T \vert \vert E^\lambda \vert \vert  J_i T_i M_T\rangle
=\nonumber \\
{{\delta _{T_fT_i}\langle J_f T_f \vert \vert
\vert E^{\lambda 0}\vert \vert \vert  J_i T_i \rangle +
\langle J_f T_f \vert \vert \vert E^{\lambda 1}\vert \vert \vert
J_i T_i \rangle
\langle T_i M_T 1 0 \vert  T_f M_T\rangle }\over {(2T_f+1)^{1/2}}}
\end{eqnarray}
where
\begin{equation}
\label{sm8}
M_T = {{Z-N}\over 2} \quad ,
\end{equation}
$Z$ and $N$ being the proton and neutron numbers of the nucleus under
consideration. In (\ref{sm7}) and elsewhere a triple bar denotes a matrix
element which is reduced in both spin and isospin spaces.

As is evident from (\ref{sm6}), the isoscalar part of the $E1$ operator
is proportional to the center-of-mass vector $\vec{R}$. Thus the $E^{10}$
operator connects a non-spurious natural parity state to a spurious one,
which belongs to the space of the unnatural parity states. As a consequence
only the isovector part of the $E1$ operator is considered in determining
$E1$ matrix elements (\ref{sm7}). However, as discussed below, the isoscalar
part of the $E1$ operator is used in the elimination of the spurious
unnatural parity states.

As described in sect. 2.1, to determine the matrix elements of $V_{coup}$
and $V_{pol}$ one needs to determine the wave functions $\vert f\rangle$ and
$\vert i\rangle$ of the final and initial states, respectively. In addition, as
eqs (\ref{df5}) and (\ref{df7}) imply, one needs to determine all states
$\vert n\rangle$ which are connected to both $\vert i\rangle$ and
$\vert f\rangle$ by the
$E1$ operator. The determination of the initial and final states presents
no problem. Their wave functions are obtained by straightforward
diagonalization of the hamiltonian in the space of $(0+2)\hbar\omega$
excitations. Moreover, since the initial and final states both belong to the
low-lying part of the spectrum, one needs to determine only few of the
eigenvalues, the lowest in energy, of the hamiltonian matrix.

On the other hand there are problems with the determination of the
states $\vert n\rangle$, discussed above.
The space of their calculation consists
of the $(1+3)\hbar\omega$ excitations and, consequently, it has a large
dimension (over 3000 for some states of $^{11}$Li). In addition the
relatively few of the states $\vert n\rangle$ which
are strongly connected by the $E1$ operator to the $\vert i\rangle$ and
$\vert f\rangle$
states need not necessarily be among the lowest in energy. Thus, in principle,
one needs to perform a full diagonalization of the hamiltonian matrix in the
space of $(1+3)\hbar\omega$ excitations. Apart from the technical difficulties,
this solution has the disadvantage that most of the eigenstates
$\vert n\rangle$
produced will couple only weakly to the initial and final states.

To face the above problem, we have adopted the BAGEL approach of Skouras and
M\"{u}ther$^{38}$) which is suitable for selecting specific eigenstates of
a hamiltonian matrix of large dimension. We describe below the manner in
which this approach was applied to the current problem.

We consider the general case where the states
$\vert i\rangle$ and $\vert f\rangle$
are different. Starting from these we can generate four states, to be denoted
by $\vert \phi _i^{(0)} \rangle \quad (i=1,..,4)$, by requiring them to
satisfy the following equations:
\begin{equation}
\label{bag1}
\langle \phi _i ^{(0)}\vert \phi _j ^{(0)} \rangle = \delta _{ij} \; ,
\quad i,j =1,...,4 \quad ,
\end{equation}
\begin{eqnarray}
\label{bag2}
E^{10}\vert i\rangle & = & a_{11}\vert \phi _1 ^{(0)}\rangle \nonumber \\
E^{10}\vert f\rangle & = & a_{21}\vert \phi _1 ^{(0)}\rangle + a_{22}
\vert \phi _2 ^{(0)}
\rangle \nonumber \\
E^{11}\vert i\rangle & = & a_{31}\vert \phi _1 ^{(0)}\rangle + a_{32}
\vert \phi _2 ^{(0)}
\rangle +a_{33}\vert \phi _3 ^{(0)}\rangle \nonumber \\
E^{11}\vert f\rangle & = & a_{41}\vert \phi _1 ^{(0)}\rangle + a_{42}
\vert \phi _2 ^{(0)}
\rangle +a_{43}\vert \phi _3 ^{(0)}\rangle +a_{44}\vert \phi _4 ^{(0)}\rangle
\end{eqnarray}
{}From the above discussion regarding the operator $E^{10}$ it follows that the
first two states $\vert \phi _1 ^{(0)}\rangle$ and
$\vert \phi _2 ^{(0)}\rangle$ are
spurious. On the other hand, the other two states
$\vert \phi _3 ^{(0)}\rangle$ and
$\vert \phi _4 ^{(0)}\rangle$ have the following properties:
i) they are orthogonal to
the spurious states generated by the isoscalar $E1$ operator and ii) they are
the only physical states in the $(1+3)\hbar\omega$ space which are connected
by the isovector $E1$ operator to both the initial and final states. In the
lowest order approximation of the BAGEL scheme the
four vectors $\vert \phi _i^{(0)}
\rangle \quad (i=1,..,4)$ define a suitable basis
for the diagonalization of the hamiltonian. Because of the construction of
this operator, the results of the diagonalization retain the separation of
spurious and non-spurious states. Thus the diagonalization provides the
two physical eigenstates $\vert n\rangle$ , those with zero isoscalar $E1$
strength, which are to be used in (\ref{df5}) and (\ref{df7}).

A better approximation to the above procedure can be obtained$^{38}$) by
treating $\vert \phi _i^{(0)} \rangle
\quad (i=1,..,4)$ as initial vectors in the
Lanczos diagonalization method$^{39}$). Thus four more vectors, to be denoted
by $\vert \phi _i^{(1)} \rangle
\quad (i=1,..,4)$ can be constructed by requiring
them
to satisfy the equations:
\begin{equation}
\label{bag3}
\langle \phi ^{(1)} _i\vert \phi ^{(1)} _j  \rangle =
\delta _{ij} \; , \quad
\langle \phi ^{(1)} _i\vert \phi ^{(0)} _j  \rangle = 0
\quad i,j =1,4 \quad ,
\end{equation}
\begin{equation}
\label{bag4}
H\vert \phi_i ^{(0)}\rangle = \sum _{j=1} ^4 b_{ij} ^{(0)}
\vert \phi ^{(0)} _j\rangle
+ \sum _{j=1} ^i c_{ij} ^{(0)}\vert \phi_j ^{(1)}\rangle \; , \quad
i=1,2,3,4 \quad .
\end{equation}
As is evident from (\ref{bag3}) the four new vectors $\vert \phi_i ^{(1)}
\rangle
\quad (i=1,4)$ carry no $E1$ strength. However, as (\ref{bag4}) implies,
these are the only four vectors that are connected with
$\vert \phi_i ^{(0)}\rangle$
through the hamiltonian. Thus diagonalization of $H$ in the eight-
dimensional space of
$\vert \phi_i ^{(0)}\rangle$ and
$\vert \phi_i ^{(1)}\rangle \quad (i=1,4)$ will result
in a better determination of the eigenvalues $E_n$ in (\ref{df5}) and
(\ref{df7}) while the $E1$ strength will be shared among the eigenvectors.
The new diagonalization again preserves the separation of spurious and
non-spurious states while the eigenvectors corresponding to physical
states are again characterized by having no isoscalar $E1$ strength.

The above procedure can be continued in the manner indicated by (\ref{bag3})
and (\ref{bag4}) by considering additional iterations where at each
iteration four additional basis vectors are introduced. However these new
vectors interact directly only with the vectors of the previous iteration
-in the basis of $\vert \phi_i ^{(k)}\rangle$ the hamiltonian matrix is
tridiagonal$^{38}$)- and thus only a small number of iterations is required
to generate the more important states $\vert n\rangle$ in (\ref{df5}) and
(\ref{df7}). Some examples of the convergence of expression (\ref{df7}) with
respect to the number of iterations are discussed in section 3.

\subsection{Two-body interaction and oscillator parameter}

In this section we discuss the determination of the two-body interaction
$G$ that appears in eq. (\ref{sm1}) and also the manner in which we selected
the value of the oscillator parameter $b =(\hbar/m\omega)^{1/2}$ which is used
to define the basis of single-particle states as presented in (\ref{sm3}).

The matrix elements of $G$ have been determined by solving the Bethe-Goldstone
equation
\begin{equation}
\label{muth1}
G= {\cal V} + {\cal V} {Q \over {E_s -QtQ}} G
\end{equation}
directly in the basis of harmonic oscillator states$^{40}$). For
the starting energy $E_s$ a constant value of -30 MeV has been adopted while
the Pauli operator $Q$ was defined to exclude any intermediate two-particle
configuration with one nucleon in a $0s$ or both nucleons in a valence
configuration, which is taken into account in our shell-model calculation.
In eq. (\ref{muth1}) the energy of the intermediate states is determined
considering only the effects of the kinetic energy operator $t$.

For the bare NN interaction ${\cal V}$ in (\ref{muth1}) we have considered
two versions of the Bonn OBE potential$^{31}$).
The parameters of these potentials have been adjusted to fit the NN scattering
phase shifts by solving the Thompson equation. The two potentials are denoted
as A and C in table A.2 of $^{31}$) and they mainly differ in the strength of
the tensor component. The potential C contains a moderate tensor component,
which yields a D-state probability for the deuteron of $P_D = 5.53$\%, while
potential A contains an even weaker tensor component which results in
$P_D=4.47$\%. In the following we shall denote the G-matrices corresponding
to potentials A and C by $G^A$ and $G^C$, respectively.

The above potentials A and C have been used in Dirac-Brueckner-Hartree-Fock
calculations for light nuclei$^{41}$). Such calculations have shown that
the Dirac spinors for nucleons in nuclear medium are substantially different
from those of free nucleons. The ratio of large to small components for the
Dirac spinors in the medium may be described in terms of an effective mass
$m^\ast$. It has been shown$^{32,41}$) that the value of $m^\ast = 630 MeV$ is
a reasonable choice for light nuclei. In our calculation we use this value
of $m^\ast$ as well as the value $m^\ast =938 MeV$ of the free nucleon (which
means that a change of the Dirac spinors in the medium is ignored) and we
shall distinguish the corresponding G-matrices by $G_m$ and $G$, respectively.

The other parameter that enters our calculation is the oscillator parameter
$b$. This is treated as a variational
parameter and we adopt the value of $b$ for which the binding energy of each
nucleus is a minimum. This procedure has been repeated for each G-matrix
considered in the calculation.

\section{Results of the calculation}

As outlined in sect. 2.3, in our calculation we have considered four types
of two-body matrix elements. These correspond to using as ${\cal V}$ in eq.
(\ref{muth1}) versions A and C of the Bonn potential$^{31}$) and values of
938 MeV and 630 MeV for the parameter $m^\ast$. In the following we shall
distinguish the results corresponding to these four sets by $G^A$, $G^C$,
$G^A _m$ and $G^C _m$. It should be noticed that the use of an
effective mass implies that the structure of Dirac spinors, which are
used to evaluate the matrix elements of the OBE potential, are
modified. However, once the determination of the two-body matrix
elements is accomplished, the remaining part of the structure calculation is
performed in a non-relativistic manner. This implies that we do not
consider the effects of $m^\star$ on the kinetic energy operator
in (\ref{sm1}).

All the G-matrices described above were determined for the values of 1.6, 1.8,
2.0 and 2.1 fm for the oscillator parameter $b$.  Thus, altogether, 16 sets
of two-body matrix elements were determined and the properties of the low-
lying states of $^6$Li and $^7$Li were calculated for all these interactions.
Since a calculation of $^{11}$Li in the large model space under consideration
requires a large amount of computer time a more restrictive selection has
been made for this isotope.

Table 1 shows the dependence of the binding energy on the interaction and the
oscillator parameter. The experimental values for this quantity are$^{42}$)
-31.99, -39.25 and -45.54 MeV for $^6$Li, $^7$Li and $^{11}$Li, respectively.
As seen in this table, all interactions considered in the calculation greatly
underestimate the binding energies of the Li isotopes. In general more binding
energy could be obtained by increasing the value for the starting energy
$E_s$ in the Bethe-Goldstone equation (\ref{muth1}). This would imply,
that one would treat the starting energy as a free parameter.
It has been our aim, however, to show how far one can reproduce the
empirical data without adjustment of parameter. Therefore we have kept
the value of $E_{s}$ at -30 MeV, which is a reasonable average of
values determined in a self-consistent scheme.

At first sight
there seems to be a discrepancy between the severe underbinding obtained
in the present investigation of Li isotopes and the very encouraging
results on the binding energy of $^{16}$O obtained by Skouras
et al~$^{38}$)
also employing OBE potentials and accounting for 2 $\hbar\omega$
configurations. One has to keep in mind, however, that the OBE
potentials employed in the present investigation are different from
those used in reference 38 although both are defined by
Machleidt$^{31}$). As we have already mentioned above the OBE potentials
used here have been obtained by solving the Thomson equation, while
those used before$^{38}$) employ the Blankenbecler Sugar equation for the
NN scattering equation. Furthermore the results presented here are
obtained from a diagonalisation in a limited space of configurations.
This emplies that the calculated energies contain effects of unlinked
diagrams, which yield repulsion, as compared to the effective operator
approach discussed in ref. 38.

The differences obtained
for the various interactions can be understood as follows: (i) The OBE
potential
with a weaker tensor component (A) yields more binding energy than the one
with a stronger tensor component (C). This is a general feature of BHF
calculations for closed shell nuclei, employing phase-shift equivalent
potentials$^{41}$), which obviously is valid also for binding energies
obtained in shell-model calculations for open shell nuclei. (ii) The
modification of the Dirac spinors in the medium contained in $G_m$ reduces
the calculated binding energy. This is also a feature already
present in DBHF calculations and which can be understood as a reduction of
the attractive components due the exchange of a scalar meson ($\sigma$).
(iii) A smaller calculated binding energy is accompanied by a larger value
of the oscillator parameter $b$ for which the minimum is obtained.
In the case of $^6$Li and $^7$Li one observes that the minimum value for
$G^A$ occurs for $b=1.8 \; fm$ although an almost equal value is obtained for
$b=1.6 \; fm$. Thus a proper variational calculation using $G^A$ interaction
most certainly will find the minimum between the above
two values of the oscillator parameter. A different behavior is observed with
the other three interactions where, as the results of table 1 indicate, the
minima are shifted to larger oscillator values. This is particularly evident
in the case of the $G^C _m$ results. This behavior is similar to features
of DBHF calculations for $^{16}$O, in which one finds that a larger tensor
force and the inclusion of Dirac effects increase the calculated radii.

Considering the results of table 1 as those of a restricted variational
calculation we obtain a natural selection for the oscillator parameter to be
considered in the rest of the calculation.
Thus for investigations on $^6$Li and $^7$Li
employing $G^A$, $G^C$ or $G^A_m$ interactions we adopt $b = 1.8 \; fm$,
while for the $G^C _m$  interaction we consider $b =2.0 \; fm$. The
corresponding values for $^{11}$Li are $b= 2.0 \; fm$ and $b = 2.1 \; fm$
for $G^A _m$ and $G^C _m$ interactions, respectively.

Fig. 1 shows the experimental and theoretical spectra of the low-lying positive
parity states of $^6$Li. The theoretical predictions on the electromagnetic
properties of this nucleus, determined for all interactions discussed above,
are compared to the experimental data$^{43}$) in table 2.

As may be seen in fig. 1, the excitation energies of the $^6$Li states with
isospin $T=0$ are rather insensitive on the interaction used and show a good
agreement with the experimental data, whereas the
position of the states with isospin $T=1$ relative to the $T=0$ states changes
quite drastically with the choice of the interaction.
This behavior is particularly evident in
the case  of the $J =0, \; T=1$ state where one obtains excitation energies
which differ by almost 2 MeV. The excitation energy of this state is
underestimated using $G^A$ and the agreement with the experiment becomes
worse using $G^C$, $G^A_m$ and $G^C_m$.

The four interactions produce quite similar results for most of the
electromagnetic observables displayed in table 2. These results are also
in reasonable agreement with the experimental data, bearing in mind
that in our calculation we consider bare electromagnetic operators.
One, however, observes
that the $G^C _m$ calculation predicts $r_{ms}$ and $B(E2)$ values which are
considerably larger than those obtained with the other interactions. This
behavior can only partly be attributed to the larger value of the
oscillator parameters $b$ used in the $G^C _m$ calculation. Therefore we
conclude that the increase in the calculated radii as we go from interaction
$G^A$ to $G^C_m$ describes the effect already observed in DBHF calculations
for closed shell nuclei (see discussion above).

Fig. 2 shows the experimental and theoretical spectra of the low-lying negative
parity states of $^7$Li. Unlike the case of $^6$Li, discussed above, the four
interactions considered in the calculation produce spectra which are quite
similar to each other and also with the experimental one. It seems that the
differences observed in the $T=1$ spectrum as compared to the $T=0$ states
of $^6$Li are getting less important for systems with more valence nucleons.
In particular, the
spectra obtained with the G-matrices derived from potential A ,
independent of the choice of the parameter $m^\ast$, are in close
agreement with the data.

To calculate the dipole polarizability
for $^7$Li one needs to calculate the unnatural parity states which are
connected by the $E1$ operator to both the ground and first excited states
of this nucleus. These states have spins $1/2^+$, $3/2^+$ and $5/2^+$ while
their isospin can be either 1/2 or 3/2. As discussed in sect. 2.2, these
states  are determined in the space of $(1+3)\hbar\omega$ excitations using
the BAGEL approach. This approach selectively determines the states which
have strong $E1$ coupling to the ground and first excited states. The results
of this calculation are shown in fig. 3 for the $T=1/2$ states and fig. 4
for the $T=3/2$. The results shown in figs. 3,4 show both the energy position
as well as the $B(E1,n\to i)$ strength from various unnatural parity states
$n$ to the ground state ($i=3/2^-$) or first excited state ($i=1/2^-$)
and correspond to the $G^C _m$ interaction. As discussed below, for this
interaction one obtains the largest, in magnitude, values of the $\tau _{if}$
tensors.

As figs. 3 and 4 show, the $E1$ strength is distributed over a large number of
states in the energy range of 10 to about 70 MeV. One may also observe a spin
dependence in the distribution. For example, in the case of the $1/2^+$ states
about 40\% of the strength is concentrated in the lowest state at about 9.5
MeV, while in the case of the $3/2^+$ states the strength is distributed over
many states and one has to extend to about 25 MeV to exhaust 50\% of the total
strength.
Moreover, one observes that there is similarity between $T=1/2$ and $T=3/2$
distributions although the latter is shifted by about 3 MeV.

In table 3 we list the theoretical predictions on the electromagnetic
properties of $^7$Li for all interactions considered in the calculation. Table
3 also includes the predictions of the calculation on the polarizability
terms $P$, $\tau _{11}$ and $\tau _{12}$. One should remark at this point
that with the existing data on $^7$Li it is not possible to obtain an
estimate for $P$~$^7$) and thus the theoretical predictions for this quantity
cannot be yet be compared with experiment. One should also observe that the
sign of $\tau _{12}$ is arbitrary since, as is evident from eq. (\ref{df7}),
it depends on the relative sign of the initial and final states which cannot
be determined uniquely. The values of $\tau _{12}$ listed in table 3
have been determined according to the convention of Barker and Woods$^{13}$),
i.e by assuming that $\langle 1 \vert \vert E2 \vert \vert 2\rangle$ is
positive.

As may be seen in table 3, the four calculations produce very similar values
for the magnetic moment of the ground state as well as for the $B(M1)$
corresponding to the deexcitation of the first excited state. On the other
hand, one observes a dependence of the $r_{ms}$, the $E2$ matrix elements
and the dipole polarizability terms on the interaction employed.
This behavior is more pronounced in the $G^C _m$ results, but one
should remember that these were obtained with a larger $b$ value. Generally,
the $G^C _m$ results are in closer agreement with experiment than those
obtained with the other three interactions.

The most sensitive quantities, of those listed in table 3, are the
polarizability
tensors $\tau _{11}$, $\tau _{12}$ and the monopole term $P$ since their
determination involves
both natural and unnatural parity states. The energy spectra of the latter
states have been determined in the BAGEL approach. Table 4 presents the
dependence of the polarizability quantities on the number of BAGEL iterations.
The results shown in table 4 correspond to the $G^C _m$ interaction but a
similar behavior was observed for the other interactions. As this table shows,
there is a very fast convergence of $\tau _{11}$, $\tau _{12}$ and $P$ with
respect to the number of iterations. In the actual calculation
of these quantities we considered 25 iterations but one observes
from table 4 that after
the fifth iteration the change of results is insignificant.

As eq. (\ref{df5}) indicates, the contribution of the various $J,T$
unnatural parity states adds up coherently in the case of the monopole
polarizability $P$. On the other hand, as is evident from (\ref{df7}),
there is a dependence of the $\tau _{if}$ tensors both on the spin as well
as on the $E1$ strength of the contributing unnatural parity states.
This dependence is examined in table 5 for both $G^A _m$ and $G^ C _m$
interactions.
As table 5 shows there is a coherent contribution from all $J,T$ states to
$\tau _{12}$ with the $J =1/2^+$ states contributing more. In contrast there
are cancellations in the various contributions to $\tau _{11}$. It is
interesting to observe that for a $J_i =3/2$ ground state, as is the
case of $^{7}$Li and $^{11}$Li, $\tau_{11}$ is given by
\begin{equation}
\label{res1}
\tau_{11} =-{{32\pi} \over 9} \sqrt {10 \over 3} \sum _n
(-1)^{J_n+3/2} W(11{3\over 2}{3\over 2};2J_n)
{B(E1; \, n\to {3\over 2}) \over (E_n -E_i)}
\end{equation}
with the Racah coefficients being positive for all possible $J_n$ values.
Thus the cancellations in $\tau_ {11}$ are entirely due to the phase
factor in (\ref{res1}).

We conclude our study of $^7$Li by examining the effects of configuration space
on the properties of this nucleus. To study these effects we made additional
calculations in the space of $0\hbar\omega$ configurations for the natural
parity states and $1\hbar\omega$ for the unnatural parity ones. In fig. 5 we
compare the low-lying energy spectrum obtained in the restricted space
with that of the extended space  while table 6 lists the remaining
properties of the natural parity states. As the results shown in both fig. 5
and table 6 indicate, there is a drastic improvement in the theoretical
predictions when one extends the model space. One, of course, should note that
the restricted space results could also be improved by a renormalization of
the operators$^{44}$).

The polarizability quantities $\tau _{12}$, $\tau _{11}$ and $P$ have also been
calculated for different choices of model spaces and the results are shown in
table 7. In one calculation, termed ``small'' in the table, the natural parity
states have been calculated in the $0\hbar\omega$ space and the unnatural in
the $1\hbar\omega$ space. In another calculation, termed ``medium'' in table 7,
the space of the natural space is extended to $(0+2)\hbar\omega$ excitations,
while the unnatural parity states are again determined in the
$1\hbar\omega$
space. Finally, ``large'' in table 7 denotes the results obtained in the
complete space used in this calculation.

As may be seen in table 7, the monopole polarizability $P$ obtains its
maximum value in the small space. On the other hand, the values obtained in
this space for the two quadrupole tensors are about 50\% in magnitude of
those obtained in the complete space. The smallest values in magnitude are
obtained for all three quantities in the medium space calculation. The
reason for this behavior is twofold: i) the coupling between $2\hbar\omega$
and $1\hbar\omega$ configurations is weak and ii) there is a considerable
increase in the energy denominators in eqs (\ref{df5}) and (\ref{df7}) as can
be deduced from the increase of binding energies listed in table 6.

The results in both tables 6 and 7 clearly suggest the importance
of high-lying
configurations which affect both the properties of the low-lying states, as
well as the polarizability effects. Extrapolating this behavior one expects
a further improvement in the shell-model results if even higher
configurations,
like $4\hbar\omega$ for the natural parity states and $5\hbar\omega$
for the unnatural parity ones could be included. Such an enlargement of the
model space could most probably make the predictions of the shell-model
similar to those of the cluster model.

The nucleus $^{11}$Li is known$^{15}$) to be a loosely bound system
with a very
small two-neutron separation energy. Therefore, the need for considering a
very large shell-model space would be more pronounced for this nucleus
than for
$^7$Li. Hence, we do not expect our parameter free calculation to account for
all the properties of $^{11}$Li despite the fact that we employ
a space in which
all configurations up to $3\hbar\omega$ excitation are included.

{}From the available experimental data on $^{11}$Li, one knows that this
nucleus
has a very large $r_{ms}$ value of $3.16 \pm 0.11 \; fm$~$^{17}$) and in
addition there is evidence that the first excited state
at 1.2 MeV has positive parity$^{45}$). From the $^7$Li investigation,
described above, we know that
such effects are best described in our model if one uses the $G^A _m$ and
$G^C _m$ interactions. Therefore, the results to be discussed below were
obtained with the use of these two interactions.

Fig. 6 shows the predictions of our calculation on the low-energy spectrum of
$^{11}$Li. As this figure shows the calculation predicts that the first
excited state of $^{11}$Li is a $1/2-$ state followed by a series of positive
parity states the lowest of which, a $3/2+$, appears at about 4 MeV.
As discussed above, one expects that the results shown in fig. 6 could change
considerably by the inclusion of higher configurations in the model space.

Fig. 7 shows the energy position and the distribution of $E1$ transition
strengths to the ground state for the $T=5/2$ positive parity states of
$^{11}$Li. The corresponding information for the $T=7/2$ states is shown in
fig. 8. The results shown in figs. 7 and 8 have been obtained using the
$G^C _m$ interaction which, as evidenced from fig. 6, produces lower
excitation energies for the unnatural parity states.
A comparison of the distributions shown in figs. 3 and 7 shows a
significant difference between $^7$Li and $^{11}$Li. In the latter,
particularly in the  $3/2^+$ case, one observes a low-energy component
in the
distribution. Specifically, the lowest $3/2^+$ state, predicted to
be at 3.86 MeV, carries about 4\% of the total $E1$ strength.
This feature could be interpreted to correspond to the {\it soft dipole mode}
speculated for this nucleus$^{15-16}$).

Finally, in table 8 we summarize the predictions of our calculation regarding
the ground state properties of $^{11}$Li. As this table shows the calculation
accounts satisfactorily for the $r_{ms}$
and the magnetic and quadrupole moments of this nucleus.
This is particularly true for the $r_{ms}$ value obtained with the
$G^C _m$ interaction. It is interesting also to note in table 8 that the
calculation predicts quite larger $P$ values than those obtained for $^7$Li
with the same interactions. This should be attributed to a) that a larger
$b$ value is used in the $^{11}$Li calculation and b) that the excitation
energy of the positive parity states is greatly reduced. On the other hand
the $^{11}$Li calculations predict smaller $\tau_ {11}$ values than those
obtained for $^7$Li. This behavior should be attributed to the strong
cancellation among the contributions of the various $J,T$ positive parity
states, an effect discussed above in connection with $^7$Li. It would be
of considerable interest if  values of $P$ and $\tau _{11}$ were obtained
in future experiments  on $^{11}$Li, since these will
provide a useful test of our calculation.

\section{Conclusions}

Results of shell-model calculations for the isotopes $^6$Li, $^7$Li and $^11$Li
are presented, which consider configurations within various major shells.
Realistic hamiltonians are considered, which contain the kinetic energy and
a NN interaction derived from modern OBE potentials$^{31}$). The effects
of NN short-range correlations are taken into account by solving the
Bethe-Goldstone equation for these potentials, considering a Pauli operator
which is consistent with the shell-model configurations taken into account.
No further renormalization of the hamiltonian and the operators for the
electromagnetic transitions has been made since it is the aim to account
for those long-range correlations by a sufficiently large shell-model space.
The effects due to spurious center of mass motion are considered by adding an
appropriate harmonic oscillator potential for the center of mass coordinate.
The main conclusions can be summarized as follows:
\begin{itemize}
\item
The bulk properties (binding energies, radii) calculated for the open shell
nuclei show a similar dependence on the OBE interaction used as it has been
observed in DBHF calculations for closed shell nuclei: NN interactions with
a stronger tensor component yield less binding energy as a phase-shift
equivalent potential with a weaker tensor force; the modification of the
Dirac spinors for the nucleons in the medium reduces the calculated binding
energy; a smaller binding energy is correlated with a larger value for the
radius.
\item
The calculated excitation spectra are weakly depending on the NN interaction.
Only in the case of $^{6}$Li a strong dependence of the energies for the
states with isospin $T=1$ relative to those with $T=0$ is observed.
A good agreement with the experimental data is achieved if a large model
space is considered.
\item
Also the  calculated electromagnetic properties show a good agreement with
the empirical data, keeping in mind that the present calculation does not
contain any adjustable parameter. Of course it is evident that our
predicitons are not as good as those obtained in phenomenological
studies employing parametrized
effective hamiltonians and electromagnetic operators.
In particular our results for the radii show poor agreement with the
empirical values. This is a well known problem of microscopic studies
of light nuclei, employing modern OBE potentials\cite{carlo}
\item
The results of the present investigation clearly indicate (see
table 7) that the polarizability tensors $\tau _{11}$ and
$\tau _{12}$ for $^7$Li depend strongly on the model space. The
fact that the present calculation produced values for these quantities
which are much  closer to the experimental estimates than those obtained in
previous shell-model studies$^{12-13}$) should be, therefore, mainly
attributed to the use of a larger space. Thus it appears that to
improve further the agreement with the experimental data one needs to go beyond
those model spaces considered in the present approach. Such an expansion
of the model space is currently very difficult to attempt due to
the exceedingly large number of shell-model configurations involved.
\item
The evaluation of the dipole polarizability requires a detailed information
on the $E1$ excitation of states which appear in a wide energy interval
(see figures 3,4,7 and 8). It was demonstrated
that the BAGEL approach provides a powerful tool to account for such
contributions in a very efficient and reliable way. This indicates that the
same approximation could be very useful for microscopic studies of double
$\beta$ decay or double charge exchange processes, where similar
summations over intermediate states occur.
It is important to repeat at this point that the BAGEL method corresponds
to exact diagonalization provided the number of iterations is
chosen to match the dimension of the energy matrix$^{38}$).
However, the advantage of the method is that one can terminate
the calculation once convergence of the quantity under consideration
has been obtained. It was found in the present calculation (see table 4)
that a small number of iteration is required to obtain
convergence for the polarizability tensors. On the other hand, a
larger number of iterations will most probably be required in
cases of non-collective excitations in which many states need to
be considered each participating with a small transition strength.
\end{itemize}

These investigations have been supported by the Commission of the European
Communities, the British Council in Athens and the Bundesministerium f\"ur
Forschung und Technologie (Germany, Project 06 T\"u 736). This support is
gratefully acknowledged.

\vfill\eject

\vfill\eject

\centerline{\bf Table 1}
\centerline{Dependence of the binding energies of the Li isotopes on the
oscillator parameter $b$}
\begin{center}
\begin{tabular}{ccccc}
\hline\hline
$b (fm)$ & 1.6    &   1.8     &   2.0  &  2.1   \\
\hline
         &        &   $^6$Li  &        &        \\
$G^A$    & -20.54 & -20.83    & -19.49 &        \\
$G^C$    & -14.46 & -15.85    & -15.43 &        \\
$G^A_m$  & -15.04 & -16.64    & -16.49 & -15.93 \\
$G^C_m$  &  -8.79 & -11.48    & -12.17 & -12.00 \\
\hline
         &        &           &        &        \\
         &        &   $^7$Li  &        &        \\
$G^A$    & -26.47 & -26.59    & -24.63 &        \\
$G^C$    & -18.89 & -20.51    & -19.75 &        \\
$G^A_m$  & -19.54 & -21.56    & -21.24 & -20.45 \\
$G^C_m$  & -11.77 & -15.26    & -16.03 & -15.74 \\
\hline
         &        &           &        &        \\
         &        & $^{11}$Li &        &        \\
$G^A_m$  &        &           & -22.03 & -21.68 \\
$G^C_m$  &        &           & -13.87 & -14.32 \\
\hline\hline\\[-2mm]
\end{tabular}
\end{center}

\vfill\eject

\centerline{\bf Table 2}
\centerline{Electromagnetic properties of $^6$Li determined with various
G-matrices}
\begin{center}
\begin{tabular}{ccccccc}
\hline\hline
Quantity & Units & Experiment$^a$) & $G^A$ & $G^C$ & $G^A_m$ & $G^C_m$ \\
\hline
$r_{ms}$ & fm & $2.09 \pm 0.02$ & 2.34 & 2.39 & 2.40 & 2.64 \\
$\mu$ & $\mu _\nu$ & 0.8220 & 0.8530 &  0.8453 & 0.8540 & 0.8524 \\
$Q_s$ & $efm^2$    & -0.083 & -0.302 & -0.407  & -0.099 & -0.365 \\
$B(E2;3,0\rightarrow 1,0)$& WU &$16.5\pm 1.3$ & 4.96 & 5.49 & 5.57 & 8.35 \\
$B(M1;0,1\rightarrow 1,0)$& WU &$8.62\pm 0.18$& 8.64 & 8.52 & 8.84 & 8.72 \\
$B(E2;2,0\rightarrow 1,0)$& WU &$6.8\pm 3.5 $ & 4.06 & 4.24 & 5.26 & 7.04 \\
$B(M1;2,1\rightarrow 1,0)$& WU$\times 10^{-2}$ &$8.35\pm1.5$ & 0.04 &
0.09 & 2.2 & 0.06 \\
\hline\hline
\end{tabular}
\end{center}
$^a$) Ref. $^{43}$).

\vspace{20mm}

\centerline{\bf Table 3}
\centerline{Electromagnetic properties of $^7$Li determined with various
G-matrices}
\begin{center}
\begin{tabular}{ccccccc}
\hline\hline
Quantity & Units & Experiment$^a$) & $G^A$ & $G^C$ & $G^A_m$ & $G^C_m$ \\
\hline
$r_{ms}$ & fm & $2.23 \pm 0.02$ & 2.41 & 2.45 & 2.46 & 2.71 \\
$\mu$ & $\mu _\nu$ & 3.2564 & 3.0789 & 3.0918 & 3.1270 & 3.1339 \\
$B(M1;1/2\rightarrow 3/2)$& WU &$2.75\pm 0.14$ & 2.39 & 2.36 & 2.39 & 2.38\\
$B(E2;1/2\rightarrow 3/2)$& WU &$19.7\pm 1.2$& 7.72 & 8.74 & 9.25 & 13.7 \\
$B(E2;7/2\rightarrow 3/2)$& WU & 4.3 & 3.69 & 4.01 & 5.65 & 6.15 \\
$Q_s$ & $efm^2$ &$-4.00\pm 0.06$& -2.54 & -2.70  & -2.81 & -3.39 \\
$B(E2;3/2\rightarrow 1/2)\uparrow$&$e^2 fm^4$& $7.27\pm 0.12$& 3.14 & 3.55 &
3.76 & 5.56\\
$P$          & $fm^3$ &                 &  0.095 &  0.109 &  0.110 &  0.164\\
$\tau _{11}$ & $fm^3$ & $-0.12\pm 0.07$ & -0.050 & -0.056 & -0.057 & -0.084\\
$\tau _{12}$ & $fm^3$ & $-0.148\pm 0.012$ & -0.049 & -0.055 & -0.053 & -0.080\\
\hline\hline
\end{tabular}
\end{center}
$^a$) The data for the last five quantities are from $^7$). The
rest are taken from $^{43}$).

\vfill\eject

\centerline{\bf Table 4}
\centerline{Dependence of the dipole polarisabitity
on the number of BAGEL iterations}
\begin{center}
\begin{tabular}{cccc}
\hline\hline
Iteration& $\tau _{12}$& $\tau _{11}$ & $P$\\
\hline
   1     & -0.06377   & -0.06662   & 0.13664\\
   2     & -0.07759   & -0.08138   & 0.16131\\
   3     & -0.07890   & -0.08368   & 0.16386\\
   4     & -0.08067   & -0.08490   & 0.16459\\
   5     & -0.07968   & -0.08325   & 0.16406\\
   6     & -0.07997   & -0.08367   & 0.16417\\
   7     & -0.08004   & -0.08376   & 0.16416\\
   8     & -0.08005   & -0.08377   & 0.16415\\
   9     & -0.08005   & -0.08376   & 0.16414\\
\hline
 Final   & -0.08005   & -0.08376   & 0.16413\\
\hline\hline
\end{tabular}
\end{center}

\vspace{20mm}

\centerline{\bf Table 5}
\centerline{Contribution of different $J,T$ states to $\tau _{if}$}
\begin{center}
\begin{tabular}{ccrrrr}
\hline\hline
\multicolumn{1}{c}{$T$} & \multicolumn{1}{c}{$J^{\pi}$} &
\multicolumn{2}{c}{$G^A _m$} & \multicolumn{2}{c}{$G^C _m$}\\
  &  &\multicolumn{1}{c}{$\tau _{12}$} &\multicolumn{1}{c}{$\tau _{11}$} &
\multicolumn{1}{c}{$\tau _{12}$} &\multicolumn{1}{c}{$\tau _{11}$} \\
\hline
$1/2$ & $1/2^+$ & -0.03465 & -0.05039 & -0.05220 & -0.07311 \\
      & $3/2^+$ & -0.00755 &  0.02759 & -0.01113 &  0.04062 \\
      & $5/2^+$ &          & -0.02091 &          & -0.03072 \\
$3/2$ & $1/2^+$ & -0.01038 & -0.02802 & -0.01630 & -0.04351 \\
      & $3/2^+$ & -0.00001 &  0.02958 & -0.00042 &  0.04531 \\
      & $5/2^+$ &          & -0.01489 &          & -0.02235 \\
\hline
\multicolumn{2}{c}{Total}& -0.05259 & -0.05702 & -0.08005 & -0.08376\\
\hline\hline
\end{tabular}
\end{center}

\vfill\eject

\centerline{\bf Table 6}
\centerline{Effects of configuration space on the electromagnetic properties
of the low-lying states of $^7$Li}
\begin{center}
\begin{tabular}{ccrrrrr}
\hline\hline
\multicolumn{1}{c}{Quantity} & \multicolumn{1}{c}{Units} &
\multicolumn{1}{c}{Experiment$^a$)} & \multicolumn{2}{c}{$G^A_m$} &
\multicolumn{2}{c}{$G^C_m$}\\
 & & & \multicolumn{1}{c}{$0\hbar\omega$}
&\multicolumn{1}{c}{$(0+2)\hbar\omega$}
&\multicolumn{1}{c}{$0\hbar\omega$} &\multicolumn{1}{c}{$(0+2)\hbar\omega$}\\
\hline
B.E      & MeV& 39.25 & 14.11 & 21.24 & 9.59 & 16.03 \\
$r_{ms}$ & fm & $2.23 \pm 0.02$ & 2.50 & 2.46 & 2.78 & 2.71 \\
$\mu$ & $\mu _\nu$ & 3.2564 & 3.1431 & 3.1270 & 3.1339 & 3.1375 \\
$B(M1;1/2\rightarrow 3/2)$& WU &$2.75\pm 0.14$ & 2.45 & 2.39 & 2.45 & 2.38\\
$B(E2;1/2\rightarrow 3/2)$& WU &$19.7\pm 1.2$& 4.17 & 9.25 & 6.67 & 13.7 \\
$B(E2;7/2\rightarrow 3/2)$& WU & 4.3 & 1.64 & 4.06 & 2.61 & 6.15 \\
$Q_s$ & $efm^2$ &$-4.00\pm 0.06$& -2.00 & -2.81 & -2.46 & -3.39 \\
$B(E2;3/2\rightarrow 1/2)\uparrow$ & $e^2 fm^4$ & $7.27\pm 0.12$ & 1.70 &
3.76 & 2.71 & 5.56\\
\hline\hline
\end{tabular}
\end{center}
$^a$) The data for the last two quantities are from $^7$). The
rest are taken from $^{43}$).

\vspace{20mm}

\centerline{\bf Table 7}
\centerline{Effects of configuration space on the dipole polarizability
of $^7$Li}
\begin{center}
\begin{tabular}{cccc}
\hline\hline
\multicolumn{1}{c}{Space} & \multicolumn{1}{c}{$\tau _{12}\; (fm ^3)$} &
\multicolumn{1}{c}{$\tau _{11}\; (fm^3)$} &\multicolumn{1}{c}{$P \; (fm^3)$}\\
\hline
Small$^a$)  & -0.0431 & -0.0463 & 0.197\\
Medium$^b$) & -0.0409 & -0.0432 & 0.095\\
Large$^c$)  & -0.0801 & -0.0838 & 0.164\\
\hline
Experiment$^d$) & -0.148 & -0.12 &     \\
\hline\hline
\end{tabular}
\end{center}
$^a$) $0\hbar\omega$ space for the negative parity states and
$1\hbar\omega$ for the positive parity states.\\
$^b$) $(0+2)\hbar\omega$ space for the negative parity states and
$1\hbar\omega$ for the positive parity states.\\
$^c$) $(0+2)\hbar\omega$ space for the negative parity states and
$(1+3)\hbar\omega$ for the positive parity states.\\
$^d$) Ref.$^7$).

\vfill\eject

\centerline{\bf Table 8}
\centerline{Properties of $^{11}$Li}
\begin{center}
\begin{tabular}{cccc}
\hline\hline
Quantity & Exper.$^a$) & $G^A _m$ & $G^C _m$ \\
\hline
$r_{ms}\; (fm)$     & $3.16 \pm 0.11$   & 2.81   & 2.97    \\
$\mu\; (\mu_{\nu})$ &$3.6673\pm 0.0025$ & 3.6908 & 3.6890  \\
$Q_s \;(efm^2)$     & $-3.12\pm 0.45$   & -3.48  & -3.92   \\
$\tau_ {11}\;(fm^3)$&                   &-0.0149 & -0.0624  \\
$P         \;(fm^3)$&                   & 0.238  &  0.375   \\
\hline\hline
\end{tabular}
\end{center}
$^a$) The value for $r_{ms}$ is from $^{17}$). For the other quantities
from $^{46}$).

\vfill\eject

\begin{center}
{\bf Figure Captions}
\end{center}
\vspace{5mm}
\def\ref{\par\smallskip\hangindent =1.75cm\hangafter=1}
\parindent=0pt
\ref{\bf Fig.\ 1: } Low-lying spectra of $^6$Li calculated with various
G-matrices (see text) in the space of $(0+2)\hbar\omega$ excitations.
All states have positive parity. Each level is labelled by $J,T$.
The experimental information is from Ref.~$^{43}$).
\ref{\bf Fig.\ 2: } Low-lying spectra of $^7$Li calculated with various
G-matrices (see text) in the space of $(0+2)\hbar\omega$ excitations.
All states have negative parity and $T =1/2$. Each level is labelled by $2J$.
The experimental information is from Ref.~$^{43}$).
\ref{\bf Fig.\ 3: } Distribution of the $E1$ strength among the $T=1/2$
positive
parity states of $^7$Li. Only transitions leading to the ground state $3/2^-$
and to the first excited $1/2^-$ state are shown. The accumulated strength
is normalized to $1$ for the complete strength in each channel.
\ref{\bf Fig.\ 4: } Distribution of the $E1$ strength among the $T=3/2$
positive
parity states of $^7$Li. Only transitions leading to the ground state $3/2^-$
and to the first excited $1/2^-$ state are shown.
\ref{\bf Fig.\ 5: } Low-lying spectra of $^7$Li calculated with interactions
$G^A _m$ and $G^C _m$ (see text). The spectra labelled (a) have been determined
in the $0\hbar\omega$ space, while those labelled (b) in the $(0+2)\hbar\omega$
space.
All states have negative parity and $T =1/2$. Each level is labelled by $2J$.
The experimental information is from Ref.~$^{43}$).
\ref{\bf Fig.\ 6: } Low-lying spectra of $^{11}$Li calculated with interactions
$G^A _m$ and $G^C _m$ (see text). All states have $T =5/2$. Each level is
labelled by $2J^{\pi}$.
\ref{\bf Fig.\ 7: } Distribution of the $E1$ strength among the $T=5/2$
positive
parity states of $^{11}$Li. Only transitions leading to the ground state are
shown.
\ref{\bf Fig.\ 8: } Distribution of the $E1$ strength among the $T=7/2$
positive
parity states of $^{11}$Li. Only transitions leading to the ground state are
shown.

\end{document}